\documentclass[superscriptaddress,aps,prb,twocolumn,preprintnumbers,amsmath,amssymb,colorlinks,floatfix]{revtex4-2}

\usepackage[dvipdfmx]{graphicx}
\usepackage{bm}
\usepackage{color}
\usepackage{natbib}
\usepackage{txfonts}
\usepackage{ulem}

\newcommand{\bcv}{BaCo$_2$V$_2$O$_8$}

\begin{document}

\title{Solitonic excitations in the Ising anisotropic chain \bcv\\
under large transverse magnetic field}

\author{Quentin Faure}
\affiliation{Univ.~Grenoble Alpes, CEA, IRIG\,/\,MEM\,/\,MDN, F-38000 Grenoble, France}
\affiliation{Institut N\'eel, CNRS--UGA, F-38042 Grenoble, France}
\affiliation{London Centre for Nanotechnology and Department of Physics and Astronomy, University College London, London, United Kingdom}

\author{Shintaro Takayoshi}
\email[Corresponding author.~Electronic address: ]{takayoshi@konan-u.ac.jp}
\affiliation{Department of Physics, Konan University, 658-8501 Kobe, Japan}

\author{B\'{e}atrice Grenier}
\affiliation{Univ.~Grenoble Alpes, CEA, IRIG\,/\,MEM\,/\,MDN, F-38000 Grenoble, France}

\author{Sylvain Petit}
\email[Corresponding author.~Electronic address: ]{sylvain.petit@cea.fr}
\affiliation{Laboratoire L\'eon Brillouin, CEA, CNRS, Universit\'e Paris-Saclay, CE-Saclay, F-91191 Gif-sur-Yvette, France}

\author{St\'{e}phane Raymond}
\affiliation{Univ.~Grenoble Alpes, CEA, IRIG\,/\,MEM\,/\,MDN, F-38000 Grenoble, France}

\author{Martin Boehm}
\affiliation{Institut Laue Langevin, CS 20156, F-38042 Grenoble, France}

\author{Pascal Lejay}
\affiliation{Institut N\'eel, CNRS--UGA, F-38042 Grenoble, France}

\author{Thierry Giamarchi}
\affiliation{Department of Quantum Matter Physics, University of Geneva, CH-1211 Geneva, Switzerland}

\author{Virginie Simonet}
\affiliation{Institut N\'eel, CNRS--UGA, F-38042 Grenoble, France}

\date{\today}

\begin{abstract}
We study the dynamics of the quasi-one-dimensional Ising-Heisenberg antiferromagnet \bcv\ under a transverse magnetic field. Combining inelastic neutron scattering experiments and theoretical analyses by field theories and numerical simulations, we mainly elucidate the structure of the spin excitation spectrum in the high field phase, appearing above the quantum phase transition point $\mu_{0}H_{c} \approx 10\;\mathrm{T}$. We find that it is characterized by collective solitonic excitations superimposed on a continuum. These solitons are strongly bound in pairs due to the effective staggered field induced by the nondiagonal $g$ tensor of the compound, and  are topologically different from the fractionalized spinons in the weak field region. The dynamical susceptibility numerically calculated with the infinite time-evolving block decimation method shows an excellent agreement with the measured spectra, which enables us to identify the dispersion branches with elementary excitations. The lowest energy dispersion has an incommensurate nature and has a local minimum at an irrational wave number due to the applied transverse field.
\end{abstract}
\maketitle


\section{Introduction}

Intensive efforts are currently being made to investigate materials exhibiting prominent quantum effects.~In this context, magnetic systems of low dimensionality make undeniable contributions with a host of different phases exhibiting strong quantum effects such as Bose-Einstein condensation~\cite{GiamarchiBECreview,Zapf2014}, spin solids and spin liquid phases~\cite{Savary2016,Takagi2019} with exotic excitations~\cite{Alicea2012,Nisoli2013}.

In the simplest case of a spin-1/2 Heisenberg chain with antiferromagnetic interactions, the ground state is strongly entangled, lacks long-range order, and hosts fractionalized excitations called spinons \cite{Giamarchi2004}.~Those peculiar excitations, quite different from classical spin waves, possess a topological nature and can be understood as domain walls that disrupt the N\'eel order and can be observed as a continuum in inelastic neutron scattering measurements.~Such physics has been realized and probed in many different experimental realizations ranging from chains to ladders, e.g.~in the quantum Heisenberg spin-chains KCuF$_3$ and CuSO$_4\bullet$5D$_2$O~\cite{Lake2005, Mourigal2013} or the quantum spin ladder (C$_5$H$_{12}$N)$_2$CuBr$_4$~\cite{Thielemann2009}.~

Systems with relatively small exchange constants provide a new avenue to study this physics, since applying a magnetic field becomes an efficient control parameter, very similar in spirit to a voltage gate for itinerant systems~\cite{Ward2013}, changing the position and even the nature of the excitation and pushing the system through quantum phase transitions.~In anisotropic 1D magnets, for instance, applying a uniform magnetic field along the Ising axis closes the gap to the lowest excitations in a way consistent with a Pokrovsky-Talapov transition~\cite{PokrovskyTalapov} and leads to an incommensurate phase \cite{kimura2007,kimura2008,canevet2013} with Tomonaga-Luttinger liquid (TLL) spin dynamics~\cite{faure2019,wang2018,bera2020}.~For a field perpendicular to the Ising-axis, one-dimensional magnetic systems undergo a quantum phase transition belonging to the well-known Ising universality class~\cite{Pfeuty1970}.~Few experimental realizations of the one-dimensional Ising model in transverse field were realized so far, e.g.~in the ferromagnetic spin-chain compound CoNb$_2$O$_6$~\cite{Coldea2010,Cabrera2014} or more recently in the antiferromagnetic spin-chains \bcv~\cite{Matsuda2017, faure2018, Wang2018-B} and SrCo$_2$V$_2$O$_8$~\cite{Wang2016, Cui2019, Kirill2020}.

\begin{figure}
\centering
\includegraphics[width=8.6cm]{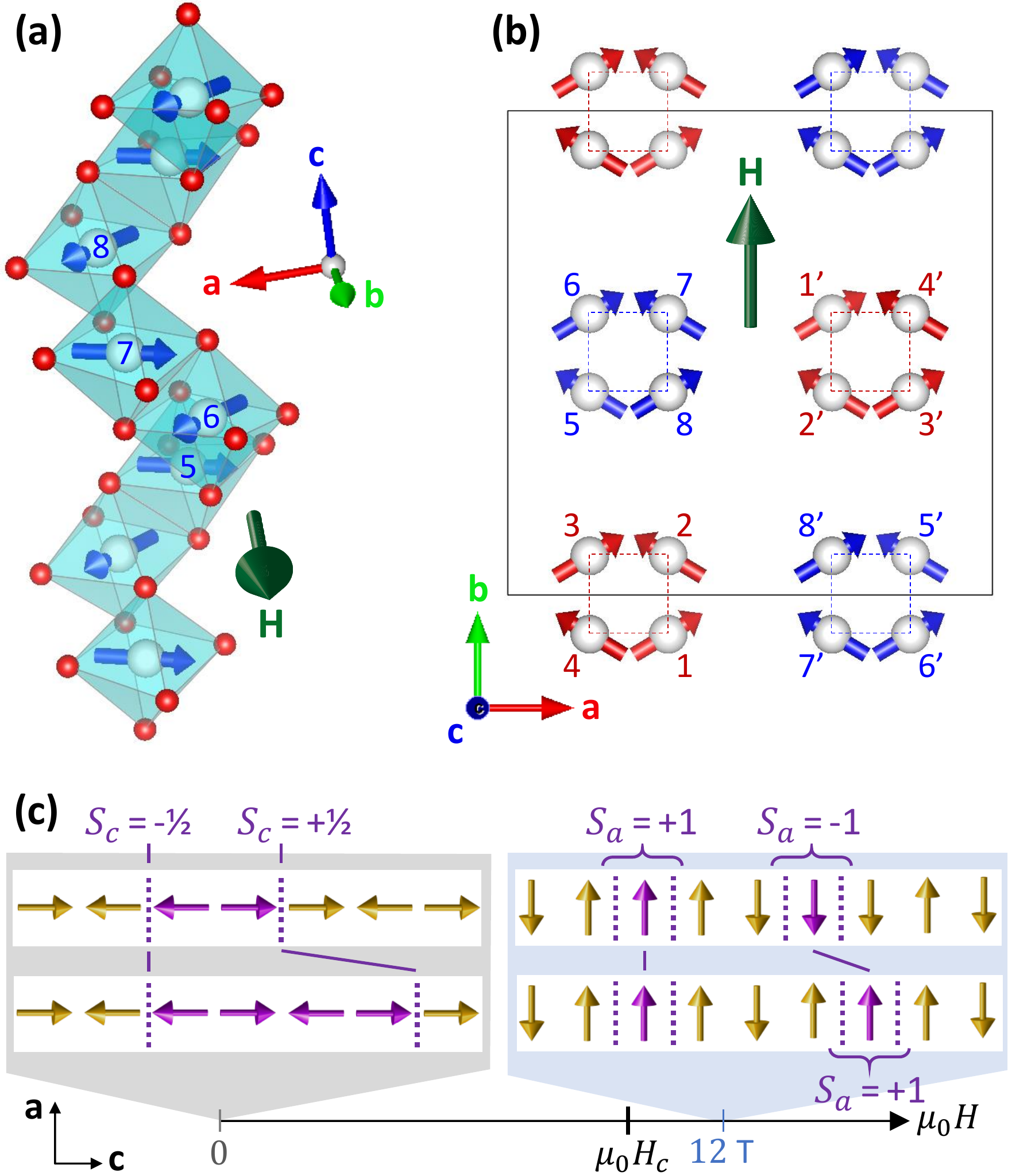}
\caption{\label{fig1} Canted antiferromagnetic structure of the Co$^{2+}$ screw chains of \bcv\ in a 12~T transverse magnetic field applied along the ${\bf b}$~axis (shown by the dark green arrow): (a) perspective view of one chain (the Co and O atoms are white and red, respectively, the CoO$_6$ octahedra are materialized in light blue, and the Co$^{2+}$ spins are represented by the blue arrows), (b) projection in the $(\mathbf{a}, \mathbf{b})$ plane of the 2 types of Co chains, with red (blue) spins for the chains having a $4_1$ (resp.~$4_3$) screw axis symmetry.~Two chains of the same type correspond to each other by the lattice centering $\frac{1}{2}(\mathbf{a}+\mathbf{b}+\mathbf{c})$.~In both panels, the labeling of the Co atoms is the same as in Refs.~\cite{canevet2013,faure2018}.~At 12 T, the $0.92\,\mu_{\mathrm{B}}$ antiferromagnetic component of the magnetic moment is aligned along the $\mathbf{a}$ axis with a field induced $0.55\,\mu_{\mathrm{B}}$ ferromagnetic component along the $\mathbf{b}$ direction.~(c) Phase diagram as a function of the applied transverse magnetic field with a sketch of the soliton excitations and their time evolution on both sides of the critical field.~At zero field, the spinon excitations carry a spin $S_c = \pm1/2$, corresponding to the topological index of the excitation, and hop by two sites along the chain axis when time evolves.~At 12 T in the high field phase, the elementary excitations carry a topological index $S_a=\pm 1$, hop by one site when time evolves, and are dual from the low field spinons~\cite{faure2018}.}
\end{figure}

Among the very rich class of materials realizing quasi-one dimensional physics, \bcv\ (see Fig.~\ref{fig1}) has indeed proven to be a specially fascinating example.~In this material, Co$^{2+}$ ions form screw chains along the ${\bf c}$ axis and carry \textit{effective} spins-1/2 coupled by antiferromagnetic exchange.~Several ingredients make it even richer: i) the chains possess a significant Ising anisotropy in the ${\bf c}$ ``chain'' direction, which is modeled by an anisotropic exchange tensor with still sizable components in the ${\bf a}$ and ${\bf b}$ directions; ii) the $g$ tensor has off diagonal staggered parts which allows to effectively apply a staggered magnetic field to the system when applying experimentally a uniform magnetic field \cite{kimura2013}; iii) the interchain dispersion is non-negligible and quite complex due to the screw nature of the chains.~As a result \bcv\ has been a perfect laboratory to tackle the exotic physics of low dimensional quantum magnets.~

Without magnetic field, \bcv\ shows a long-range N\'eel order below $T_{\mathrm{N}}=5.5$~K where the spins align antiferromagnetically along the ${\bf c}$ axis, i.e., the direction of Ising anisotropy \cite{He2005}.~In the presence of such N\'eel order the spinons, which would be free for an isolated single chain, are confined by the linear potential due to the interchain coupling, giving rise to a series of bound states, that have been observed by neutrons \cite{grenier2015} as a series of discrete excitations, in agreement with the theoretical expectations.~Similar effects have been reported in the sister compound studied in \cite{Bera2014,Bera2017}.~Note that although existing for a 3D compound, those modes, also called Zeeman ladders~\cite{ishimura1980}, are different from classical spin waves, a remarkable result which is due to the small yet sizable value of the interchain coupling.~It is worth noting that, in the case of an antiferromagnetic spin-chain, any staggered field, arising from interchain couplings as described above or from non-diagonal terms in the $g$ tensor, was shown both experimentally~\cite{Kenzelmann2004, grenier2015, Wang2015, faure2018} and analytically~\cite{Oshikawa1997, Affleck1999, Rutkevich2018} to confine spinons.

The staggered parts in the $g$ tensor are responsible for an even richer situation.~The system undergoes a transition at $\mu_0H_c \approx 10$~T that was identified~\cite{Okutani2015,faure2018,Matsuda2017}, by a combination of field theory, numerical analysis and neutron scattering experiments, as a spin-flop transition from the ${\bf c}$ to ${\bf a}$ direction.
This transition, which is in the same universality class as the celebrated transverse field Ising model one \cite{Pfeuty1970}, is characterized by different topological excitations above and below $\mu_0H_c$~\cite{faure2018} and is generally described by a dual field double sine-Gordon model~\cite{Takayoshi2018}.~The evolution of the spin correlations as measured by polarized neutrons was studied for fields up to 12~T with a special focus on the low energy modes, including their nature and polarization.

The aim of the present paper is to continue such investigations in the high field phase, above the transition, combining an inelastic neutron scattering study of the spin-spin correlations with a theoretical analysis based on the model that was introduced in~\cite{faure2018}.~Such an analysis is particularly useful in the case of \bcv\ given its complexity and the various ingredients at play, namely the spin anisotropy, the interchain coupling and the ``two'' magnetic fields (uniform and staggered) effectively applied to the system due to the nature of the $g$ tensor.~The comparison between measurements and theory allows us to precisely analyze the spectrum of the new soliton excitations in the high field phase and to thoroughly disentangle the influence of the various ingredients at play in \bcv\ on the physics of this phase.~This is especially important for the two components of the magnetic field since the uniform one is responsible for the incommensurability of the spin excitations while the staggered one is the one determining the quantum phase transition.~Although most of the properties are well in line with the model introduced in~\cite{faure2018}, some yet unexplained additional features emerge, such as a tetramerization of the spectrum.~

The plan of the paper is as follows: Section~\ref{sec:methods} provides the details of our experimental and theoretical approaches.~Section~\ref{sec:results} describes the neutron measurements of the spin-spin correlation functions and its interpretation in terms of the corresponding numerical simulations, based essentially on the infinite time evolving block decimation (iTEBD) technique~\cite{Vidal2007} with a mean-field treatment of the coupling between the chains.~We focus in particular on the identification of the collective modes and on their polarization.~Section~\ref{sec:discuss} discusses these results in the light of a simplified model introduced to better understand and highlight the important features of the dispersions.~Conclusions and perspectives are found in Sec.~\ref{sec:conclusion}.

\section{System, model, and methods}
\label{sec:methods}

In this section, we describe how to model the target material, \bcv, as a quasi-one dimensional spin system schematically represented in Fig.~\ref{fig1}.~We also explain our experimental and theoretical methods to study the microscopic mechanisms at the origin of the dynamics of the system.~

\subsection{Experimental measurements}

The \bcv\ single-crystal was grown at Institut N\'{e}el by the floating zone method~\cite{lejay2011}.~It was aligned with the ${\bf b}$ axis vertical yielding $({\bf a^\star}, {\bf c^\star)}$ as the horizontal scattering plane and placed in a cryomagnet providing a maximum uniform field of 12~T at a base temperature of 1.5~K.~Inelastic neutron scattering (INS) experiments were performed on two cold-neutron triple-axis spectrometers, ThALES and FZJ-CRG IN12, both installed at ILL (France).~On ThALES~\cite{boehm2015}, a PG(002) monochromator (resp.~analyzer) was used to select (resp.~analyze) the initial (resp.~final) wave vector of the unpolarized neutron beam.~On IN12, the spin of the incident neutrons was polarized using a cavity transmission polarizer located far upstream in the guide.~The initial wave vector was selected by a PG(002) monochromator, and both the final wavevector and neutron polarization were analyzed using a Heusler analyzer (see Ref.~\cite{Schmalzl2016} for a more detailed description of the standard polarized neutron setup on IN12).~On both spectrometers, the energy resolution was of the order of 0.15~meV and the high order harmonics were suppressed by a velocity selector.~The polarization analysis performed on IN12 uses the classical $(X,Y,Z)$ frame where the $X$ axis is aligned with the scattered wave vector ${\bf Q}$, the $Z$ axis is vertical and the $Y$ axis is perpendicular to ${\bf Q}$ and $Z$.~The strong applied vertical magnetic field (up to 12~T) restricts the polarization analysis to the so called $P_Z$ channel.~Scattering which involves a spin flip then encodes the correlations between $Y$ components of the spins, while the non-spin-flip scattering encodes the correlations between $Z$ components (on top of the nuclear scattering).~Importantly, the vertical current of the Mezei spin flipper, placed just before the monochromator on IN12, was calibrated for every used value of the incident wave vector and of the magnetic field.~The horizontal current was checked to be non sensitive to the applied field.~The flipping ratios were determined using a graphite sample: their values were ranging between 12 and 23, depending on the incident wave vector and magnetic field values.

\subsection{Model}

We consider the model Hamiltonian for \bcv\
\begin{align}
 \mathcal{H} =
  J \sum_{n,\mu} [\epsilon
   (S_{n,\mu}^{a}S_{n+1,\mu}^{a}+S_{n,\mu}^{b}S_{n+1,\mu}^{b})
   +S_{n,\mu}^c S_{n+1,\mu}^c] \nonumber\\
  -\sum_{n,\mu} \mu_0 \mu_{\mathrm{B}} \mathbf{H} \cdot \tilde{g}\mathbf{S}_{n,\mu}
  +J'\sum_{n}\sum_{\langle \nu \mu \rangle} S_{n,\mu}^{c} S_{n,\nu}^{c}
\label{eq:Hamil}
\end{align}
also used in Ref.~\cite{faure2018}.~The first term is the $XXZ$ Hamiltonian where $\mathbf{S}_{n,\mu}$ is a spin-1/2 operator, $\mu$ is the chain index, $n$ is the site index, $J=5.8\;\mathrm{meV}$ is the antiferromagnetic (AF) intrachain exchange coupling and $\epsilon=0.53$ is the magnetic anisotropy.~The second term is the Zeeman term arising from the application of the magnetic field.~$\mu_{\mathrm{B}}$ is the Bohr magneton, $\tilde{g}$ the Land\'e $g$ tensor, and $\mathbf{H}$ the external magnetic field (applied along the {\bf b} axis).~The last term describes the weak interchain coupling.~We consider here the simplest form for this coupling, namely a uniform antiferromagnetic unfrustrated nearest neighbor term.~Note that the precise nature of the interchain coupling in \bcv\ is still largely unknown with most likely more complex terms occurring due to the screw nature of the chains~\cite{Klanjsek2015}.~The interchain term taken here should thus be seen as a phenomenological term, resulting potentially from the average of several individual couplings.~The value giving at zero magnetic field the best comparison with the experimental results for the spinon confinement \cite{faure2018} is $J'=0.17$~meV.~This is the value that we take in the present paper.~

Due to tilting of the ligands of the Co$^{2+}$, the $\tilde{g}$ tensor becomes  nondiagonal~\cite{kimura2013}.~The influence of an applied uniform field along the ${\bf b}$ axis given this $\tilde{g}$ tensor is then described by
\begin{align}
 \mathbf{H} \cdot \tilde{g} \mathbf{S}_{n,\mu}
  =H \Big[g_{ba} (-1)^{n}S_{n,\mu}^{a}+g_{bb}S_{n,\mu}^{b}
   +g_{bc}\cos\Big(\pi\frac{2n-1}{4}\Big)S_{n,\mu}^{c}\Big]
\label{eq2}
\end{align}
with $g_{ba}/g_{bb}=0.40$, $g_{bc}/g_{bb}=0.14$, and $g_{bb}=2.35$.~The third term in Eq.~\eqref{eq2} is a four-site periodic field, but its effect is negligible \cite{faure2018}.

Owing to the effective staggered field [first term in Eq.~\eqref{eq2}] induced by the nondiagonal $\tilde{g}$ tensor, it is favorable for the spins to cant from the ${\bf c}$ to ${\bf a}$ direction with increasing applied field.~A quantum phase transition eventually occurs at around $\mu_0H_c \approx 10$~T, above which the spins are essentially aligned along the ${\bf a}$~axis and get progressively polarized by the uniform field in the ${\bf b}$ direction [see Figs.~\ref{fig1}(a) and \ref{fig1}(b)].~Moreover, this transition can be considered as a topological transition as it separates two phases which host different types of topological excitations.~The latter are dual from each other and well described by the double-sine Gordon model~\cite{faure2018}.
Below $\mu_0H_c$, the excitations are spinons while they become solitons carrying a topological index $S_{x(=a)}=\pm1$ along the ${\bf a}$ direction above $\mu_0H_c$.~In Ref.~\cite{faure2018}, the evolution of the spectrum under a transverse magnetic field up to 12~T was studied.~Special attention was paid to the lowest energy excitations at selected positions of the reciprocal space but the instructive full spectrum of the solitons in the high-field phase was not investigated.~In the subsequent sections of the present paper we focus on this point and investigate by inelastic neutron scattering the dispersion along the ${\bf c}$ axis of the solitonic excitations above the critical field $H_c$.

\subsection{Numerical calculations}

In the theoretical approach employed in this paper, we treat the interchain coupling in Eq.~\eqref{eq:Hamil} using a mean-field approximation
\begin{align}
 J'\sum_{n}\sum_{\langle \mu,\nu \rangle}
  S_{n,\mu}^{c} S_{n,\nu}^{c}
  \simeq J'\sum_{n,\mu}\sum_{\langle \nu \rangle_{\mu}}
  S_{n,\mu}^{c} \langle S_{n,\nu}^{c}\rangle,
\nonumber
\end{align}
which allows to reduce the quasi-1D problem to an effective one-dimensional one in the presence of a (self-consistent) staggered field.~Mean fields arising from interchain coupling with $S^{a}$ and $S^{b}$ are much smaller than the magnetic fields and are thus considered as negligible.~Therefore, the Hamiltonian is recast into the simple chain problem
\begin{align}
 \mathcal{H} &=
  J \sum_{n} [\epsilon
   (S_{n}^{a}S_{n+1}^{a}+S_{n}^{b}S_{n+1}^{b})
   +S_{n}^c S_{n+1}^c] \nonumber\\
  &-\mu_{0} \mu_{\mathrm{B}} H
   \sum_{n} \Big[g_{ba} (-1)^{n}S_{n}^{a}+g_{bb}S_{n}^{b}
   +g_{bc}\cos\Big(\pi\frac{2n-1}{4}\Big)S_{n}^{c}\Big]
\nonumber\\
  &+\tilde{h}_{c}\sum_{n}(-1)^{n} S_{n}^{c},
\end{align}
where $\tilde{h}_{c}=J'|\langle S_{n}^{c}\rangle|$ is an effective staggered field induced by the interchain coupling.~Note that the number of nearest neighbor chain site is only due to the spiral structure of $\mathrm{Co}^{2+}$ ions.

For the numerical simulations, we first obtain the ground state of the Hamiltonian by infinite density matrix renormalization group (iDMRG)~\cite{McCulloch2008}, and then calculate the retarded spin-spin correlation function:
\begin{align}
 C_{\alpha\beta}(\mathbf{r},t)
  =-i\theta_{\mathrm{s}}(t)
   \langle[S_{\mathbf{r}}^{\alpha}(t),S_{0}^{\beta}(0)]\rangle,
\nonumber
\end{align}
($\theta_{\mathrm{s}}(t)$ is the step function)
by using iTEBD~\cite{Vidal2007} with infinite boundary condition~\cite{Phien2012}.~The dynamical susceptibility is the Fourier transform of the retarded correlation function,
\begin{equation}
 \chi_{\alpha\beta}(\mathbf{Q},\omega)
  =\int_{-\infty}^{\infty}dt\sum_{\mathbf{r}}
   e^{i(\omega t-\mathbf{Q}\cdot\mathbf{r})}
   C_{\alpha\beta}(\mathbf{r},t)
\nonumber
\end{equation}
which is related to the INS spectrum $S(\mathbf{Q},\omega)$ by
\begin{align}
 S(\mathbf{Q},\omega)\propto
  \sum_{\alpha,\beta=x,y,z}
  \Big(\delta_{\alpha,\beta}
   -\frac{Q_{\alpha}Q_{\beta}}{|\mathbf{Q}|^{2}}\Big)
   S_{\alpha\beta}(\mathbf{Q},\omega),
\nonumber
\end{align}
where
$S_{\alpha\beta}(\mathbf{Q},\omega)
=\big|\mathrm{Im}\,\chi_{\alpha\beta}(\mathbf{Q},\omega)\big|$.
For the calculations of the INS spectrum (especially the lattice Fourier transform), we employ the actual positions of Co$^{2+}$ ions.~The dimension of the matrix product representation for iDMRG and iTEBD is $60$, the discrete time step is $dt/(\epsilon J)^{-1}=0.05$, and the spin-spin correlation is calculated for the time interval $0\leq dt/(\epsilon J)^{-1}\leq 60\;\mathrm{or}\;80$.

To better identify the excitations, we shall also consider a simplified model where the spins occupy the sites of a simple linear chain:
\begin{align}
 \mathcal{H} =&
  J \sum_{n} [\epsilon
   (S_{n}^{a}S_{n+1}^{a}+S_{n}^{b}S_{n+1}^{b})
   +S_{n}^c S_{n+1}^c] \nonumber\\
  &-g_{ba}\mu_{\mathrm{B}} \mu_0 H\sum_{n}(-1)^{n}S_{n}^{a}
   -g_{bb}\mu_{\mathrm{B}} \mu_0 H\sum_{n}S_{n}^{b},
\label{eq:SimpleHamil}
\end{align}

\section{Results}
\label{sec:results}

In this section, we present the results of the INS measurements and elucidate the dynamics of \bcv\ under the large transverse magnetic field by comparing the experimental data with the theory essentially based on the numerical simulation.

\subsection{Spin dynamics along the chain}

\begin{figure}
\centering
\includegraphics[width=8.5cm]{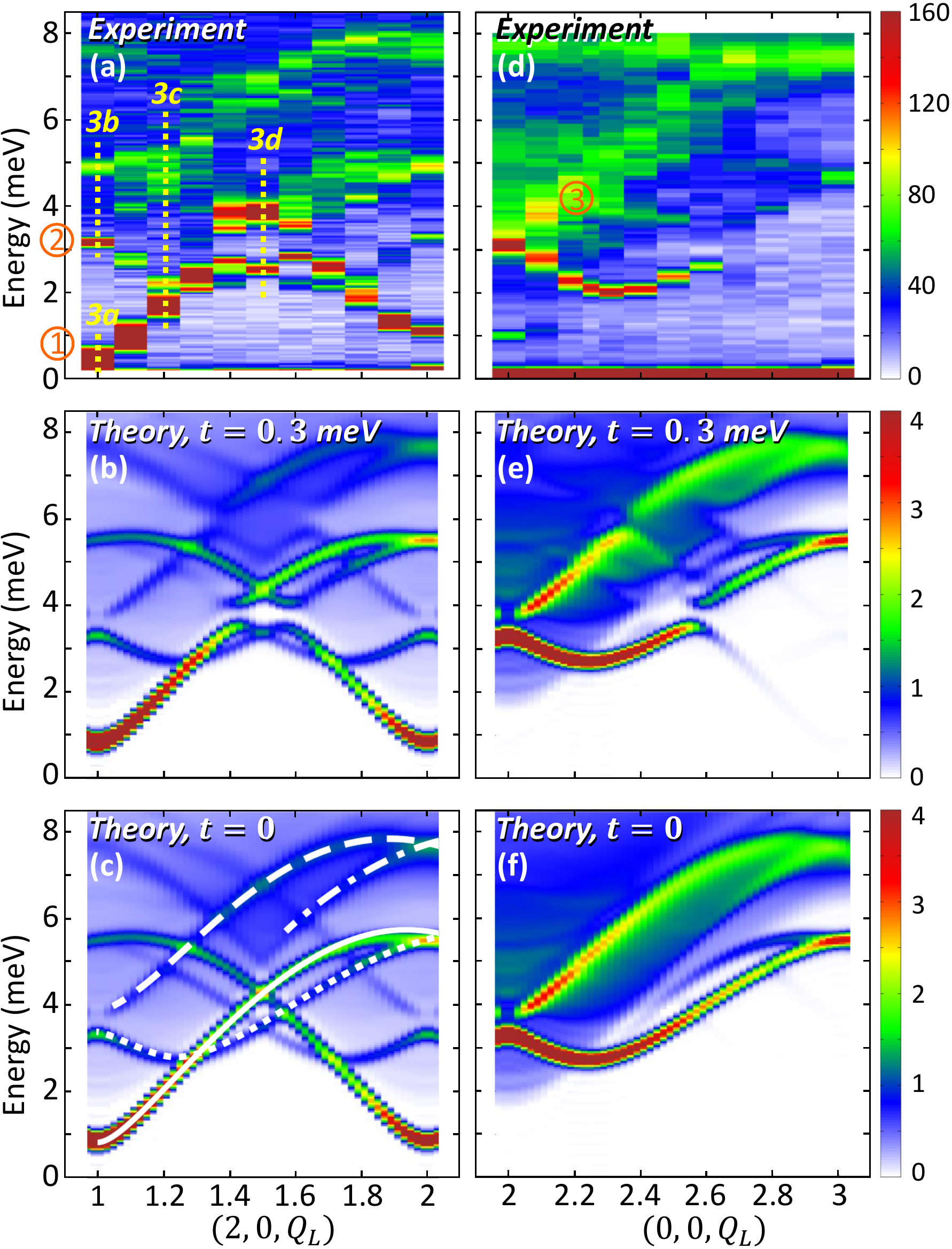}
\caption{\label{fig2} Inelastic scattering intensity maps showing the intrachain dispersion of the magnetic excitations along (a-c) $\mathbf{Q}=(2,0,Q_{L})$ and (d-f) ${\bf Q}=(0,0,Q_L)$, in a transverse field of 12 T applied along the ${\bf b}$ axis: (a,d) maps obtained experimentally on ThALES and IN12, respectively, from a series of constant-${\bf Q}$ energy scans, compared with the numerically calculated neutron scattering cross section (b,e) of a tetramerized chain with $t=0.3$~meV and (c,f) of a uniform chain ($t=0$), as explained in the text.~In panel (a), the yellow dotted lines show the energy scans presented in Figs.~\ref{fig3}(a-d) and the first three modes are labeled in panels (a) and (b) by the orange numbers.~The white lines in panel (c) correspond to the same three modes plus a fourth one, all sketched in Fig.~\ref{fig6} with the same symbols (solid, dotted, dashed, and dash-dot lines for modes \textcircled{1}, \textcircled{2}, \textcircled{3}, and \textcircled{4}, respectively).}
\end{figure}

The spectrum of the magnetic excitations measured by means of INS at $\mu_0 H=12~\mathrm{T}$ is plotted in Figs.~\ref{fig2}(a) and \ref{fig2}(d).~The experimental intensity maps are constructed from constant-$\mathbf{Q}$ energy scans taken along $(2,0,Q_L)$ and $(0,0,Q_L)$.~They were recorded with a 0.1 r.l.u.~step in $Q_L$ and an energy transfer varying between 0.2 and 8.2~meV.

As already reported in Ref.~\cite{faure2018}, the discrete Zeeman ladder spectrum characterizing the zero-field spin dynamics (see Fig.~1 in Ref.~\cite{grenier2015}) becomes less evident when increasing the uniform field, and is lost above the $\mu_0 H_c \approx 10$~T critical field which signals the quantum phase transition.~The present results at 12 T exhibit well-defined branches, drastically different from these Zeeman ladders.~Some diffuse intensity is also observed above 2 meV, which reflects underlying continua of excitations.~Note that ``spurious'' (parasitic) intensities were identified in the map of Fig.~\ref{fig2}(d) at about 1~meV around $Q_L=2$ and around 7.5~meV for $2\leq Q_L\leq 2.6$.

The low energy excitations characterized by the largest spectral weight are labeled \textcircled{1} and \textcircled{2} [see Fig.~\ref{fig2}(a)].~Leaving aside the anticrossing at ${\bf Q}=(2,0,1.5)$ discussed below, mode \textcircled{1} disperses throughout the Brillouin zone from 0.4 meV at $Q_L=1$, up to 5.5~meV at $Q_L=2$.~Mode \textcircled{2} is observed at 3.2 meV for $Q_L=1$ and disperses downwards and then upwards reaching also 5.5~meV at $Q_L=2$.~The third branch, labeled \textcircled{3}, especially visible in Fig.~\ref{fig2}(d), starts from 3.2~meV at $Q_L=2$ and disperses up to 7~meV at $Q_L=3$.~An extinction in the spectral weight, at $\approx 3$~meV, is observed for various $Q_L$ values, especially clearly around ${\bf Q}=(2,0,1.5)$ and ${\bf Q}=(0,0,2.6)$.~It is attributed to an anticrossing between modes \textcircled{1} and \textcircled{2}.

The numerically calculated INS spectra with the Hamiltonian Eq.~\eqref{eq:Hamil} in a magnetic field of $\mu_{0}H=12\;\mathrm{T}$ are displayed in Figs.~\ref{fig2}(c) and \ref{fig2}(f).~They are in good agreement with the experiment.~The well-defined excitations coexist with a continuum, especially visible in the ${\bf Q}=(0,0,Q_L)$ map with a minimum of $\approx 2\;\mathrm{meV}$ at $Q_L=2$.~These results however do not reproduce the anticrossing at the wave vector $Q_L = 1/2$.~It is instructive to translate this wave vector using the reduced reciprocal units associated to the fictitious lattice spacing $c_0=c/4$ between two neighboring Co$^{2+}$ ions along the chain direction.~The physical meaning of the anticrossing point at $\frac{2\pi}{c}Q_L=\frac{1}{2}\frac{\pi}{2c_0}$ is the loss of the four-site translational symmetry.~The original band and its shifted replica by $\frac{\pi}{2c_0}$ are coupled, yielding the opening of a gap at the crossing point.~Although the last term in Eq.~\eqref{eq2}, which is four-site periodic, may be a good candidate, the result of numerical calculations [Figs.~\ref{fig2}(c,f)] shows that its influence is negligible and cannot be the cause of the anticrossing.~To describe this anticrossing, we added a phenomenological isotropic tetramerization term
\begin{align}
 -t\sum_{n}\sum_{\mu}&
  \sqrt{2}\cos\Big(\pi\frac{2n+1}{4}\Big)\,
   \mathbf{S}_{n,\mu}\cdot\mathbf{S}_{n+1,\mu}
\label{eq:tetramer}
\end{align}
to the Hamiltonian Eq.~\eqref{eq:Hamil}.~In Figs.~\ref{fig2}(b) and \ref{fig2}(e), we display the calculated spectra for the tetramerization parameter $t=0.3\;\mathrm{meV}$.~This additional ingredient indeed accounts for the anticrossing, which is note worthily already observed at zero magnetic field~\cite{grenier2015}.~The origin of this tetramerization could not be determined on experimental grounds and remains unclear at the present stage.~It would require for instance an additional symmetry lowering with respect to the tetragonal to orthorhombic one already reported~\cite{Niesen2013}.~At least, our results indicate that the four-site periodic term is sufficient to cause the anticrossing at the proper wave vector.~Another possible origin for this kind of four-site periodicity is the complicated interchain effects arising from the spiral structure of \bcv.~More detailed investigations of the interchain interactions are beyond the scope of this paper, and are left for a future study.

\subsection{Spin polarization of the excitations}

\begin{figure}
\centering
\includegraphics[width=8.6cm]{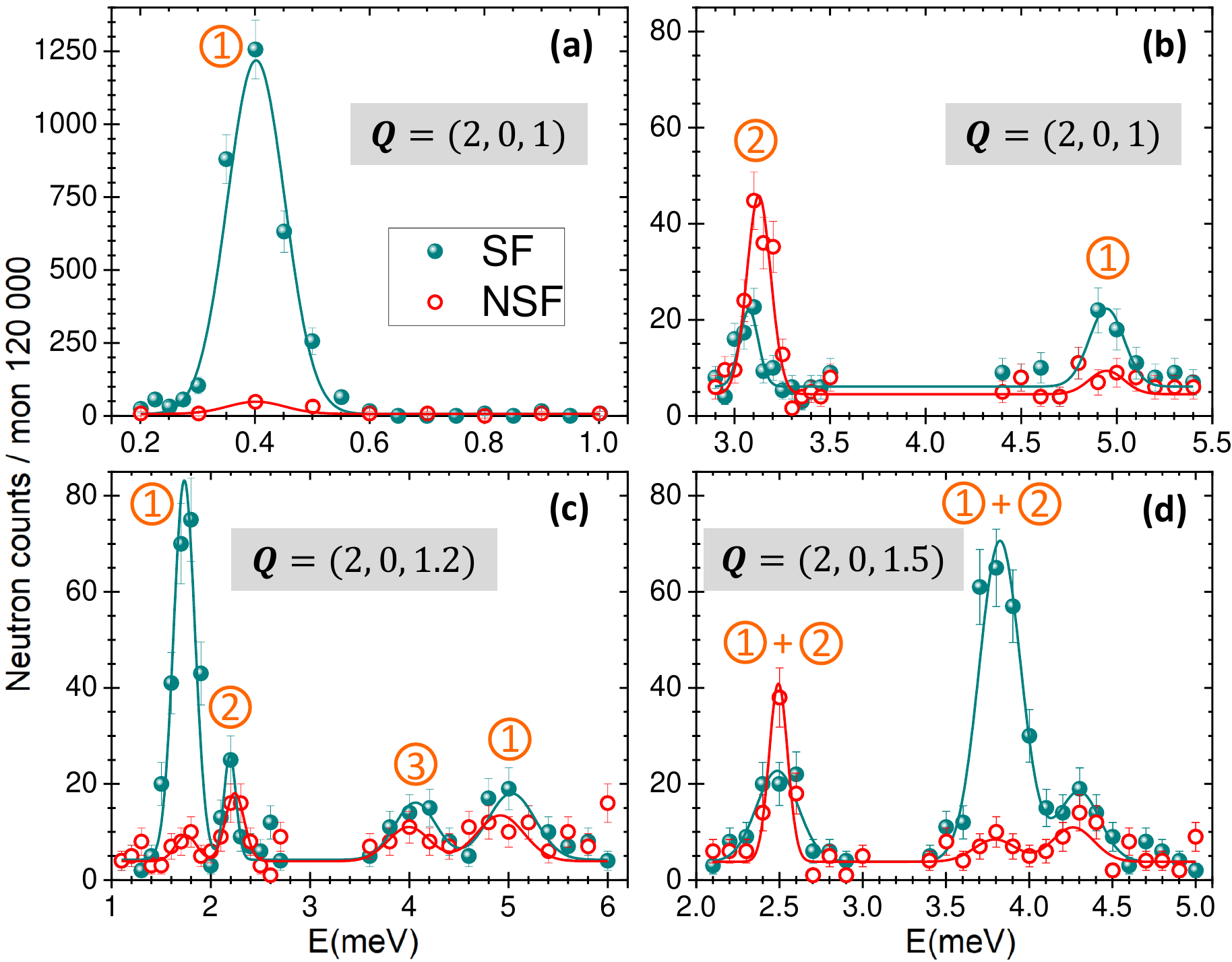}
\caption{\label{fig3} Constant$-\bf Q$ energy scans measured on IN12 using polarized neutrons in the spin-flip (SF) and non spin-flip (NSF) channels at (a,b) ${\bf Q}=(2,0,1)$, (c) ${\bf Q}=(2,0,1.2)$, and (d) ${\bf Q}=(2,0,1.5)$.~These scans are materialized by the yellow dotted lines on the map of Fig.~\ref{fig2}(a).~Note the about 16 times larger intensity scale for panel (a) as compared to panels (b-d).~The SF intensity corresponds to fluctuations occurring in the $({\bf a}, {\bf c})$ plane and perpendicular to the scattering vector ${\bf Q}$ while the NSF one corresponds to fluctuations along the ${\bf b}$ axis.}
\end{figure}

The polarization of the various excitations in the high field phase can be further studied both experimentally and numerically.~To this end, inelastic neutron scattering experiments were conducted on IN12, using polarized neutrons.~Since a strong magnetic field (12~T) is applied along the vertical ${\bf b}$ axis, the spin flip (SF) and non spin flip (NSF) channels correspond to correlations between spin components perpendicular to the scattering wave vector ${\bf Q}$ within the $({\bf a}, {\bf c}$) plane and between spin components along the ${\bf b}$ axis, respectively.~Figure~\ref{fig3} shows constant-${\bf Q}$ energy scans performed in both channels at the $Q_L$ positions and $E$ ranges spotted in Fig.~\ref{fig2}(a) by the yellow dotted lines.

The SF data allow to follow mode \textcircled{1} (green points in Figure~\ref{fig3}).~It is visible at 0.4 meV for $Q_L=1$, at 1.7~meV for $Q_L=1.2$, and distributed on both sides of the anticrossing for $Q_L=1.5$.~Hence, it is mostly polarized in the $({\bf a}, {\bf c})$ plane.~We can further deduce that its polarization is along the ${\bf c}$ axis as it is absent in the ${\bf Q}=(0,0,Q_L)$ map.~Mode \textcircled{1} is thus a transverse fluctuation, with respect to the ordered antiferromagnetic moments (along $\bf a$), and involves spin components along ${\bf c}$.~In the same $Q_L$-range, mode \textcircled{2} appears with a larger intensity in the NSF channel (red points in Fig.~\ref{fig3}) at $Q_L=1$ and 3.15 meV, showing that it is more polarized along the ${\bf b}$ axis than in the other two directions [see Figs.~\ref{fig3}(b-d)].~It becomes more mixed with the other polarizations at higher energy.~This confirms our previous understanding of these modes and definitely demonstrates their transverse character, perpendicular to the ordered moments in the high field phase~\cite{faure2018}.~Mode \textcircled{3}, visible at 4~meV in Fig.~\ref{fig3}(c), seems to present equal SF and NSF contributions, which should result from a mixture of the ${\bf a}$, ${\bf b}$ and ${\bf c}$ polarizations.

\begin{figure}
\centering
\includegraphics[width=8.5cm]{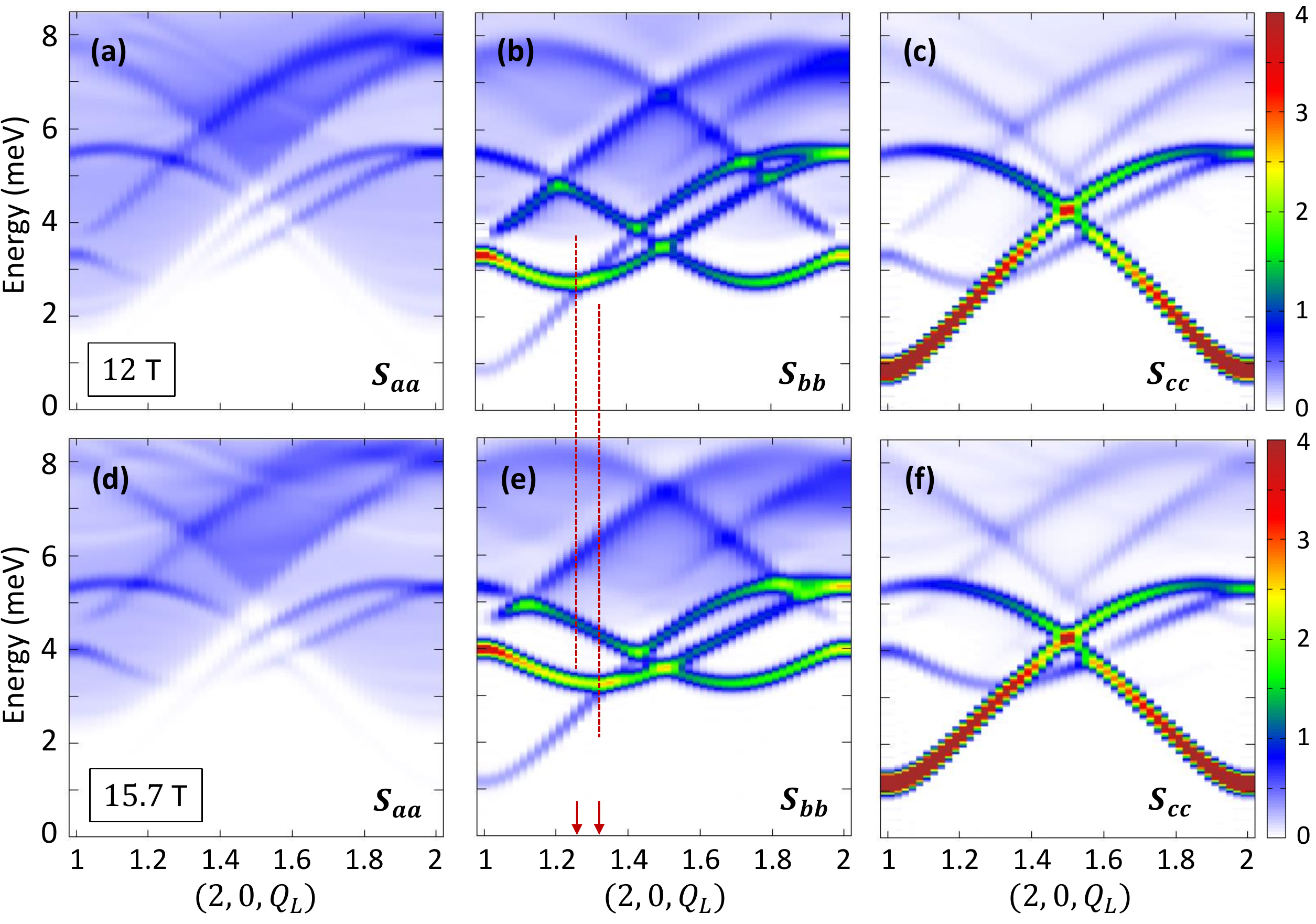}
\caption{\label{fig4} (a-c) $S_{aa}$, $S_{bb}$, and $S_{cc}$ components of the intensity color map calculated at 12~T and shown in Fig.~\ref{fig2}(c).~Panels (d-f) are the same quantities but calculated in a higher magnetic field of 15.7~T.~The dotted red lines show the shift of the incommensurate minimum, from $Q_L\simeq 1.26$ to $Q_L\simeq 1.32$, with increasing magnetic field.}
\end{figure}

These experimental determinations of the magnetic excitations polarization have been further inspected using iTEBD calculations. Figures~\ref{fig4}(a-c) show the components of dynamical structure factors calculated at 12~T for $t=0$ (i.e.~with no tetramerization) and spin components along the ${\bf a}$, ${\bf b}$ and ${\bf c}$ axes, labeled respectively $S_{aa}$, $S_{bb}$, and $S_{cc}$.~It confirms that modes \textcircled{1} and \textcircled{2} are respectively mainly polarized along the ${\bf c}$ and ${\bf b}$ axis, while modes \textcircled{3} and \textcircled{4} are polarized in the three directions ${\bf a}$, ${\bf b}$, and ${\bf c}$.~Overall, the mixing of polarization increases with the energy for all excitations.~

Note that the total spectral weight of the well-defined excitations is mostly transverse, i.e.~polarized perpendicular to ${\bf a}$.~Longitudinal spin dynamics (polarized along ${\bf a}$) is hardly observed in this 12~T magnetic phase at low energy, contrary to the zero-field phase \cite{grenier2015}.~On the other hand, a continuum of excitations is clearly visible at 12~T in the longitudinal channel with a minimum at $\approx 2$~meV.~

\subsection{Identification of the collective modes}

It is quite difficult to single out the excitation modes in the above results.~This is due to the fact that the crystalline structure of \bcv\ induces a significant complication on the spin excitation spectrum.~In particular, it is necessary to include 4 ions in the unit cell (along {\bf c}) to get a proper description of a single chain.~This redundancy, with respect to an ideal uniform chain where spins would be placed regularly along the chain, gives rise to folding effects, and more specifically to four replicas of the main dispersions, shifted along the $\mathbf{c}^{*}$ direction by one reciprocal lattice unit ($\Delta Q_L=1$).

\begin{figure}
\centering
\includegraphics[width=8.6cm]{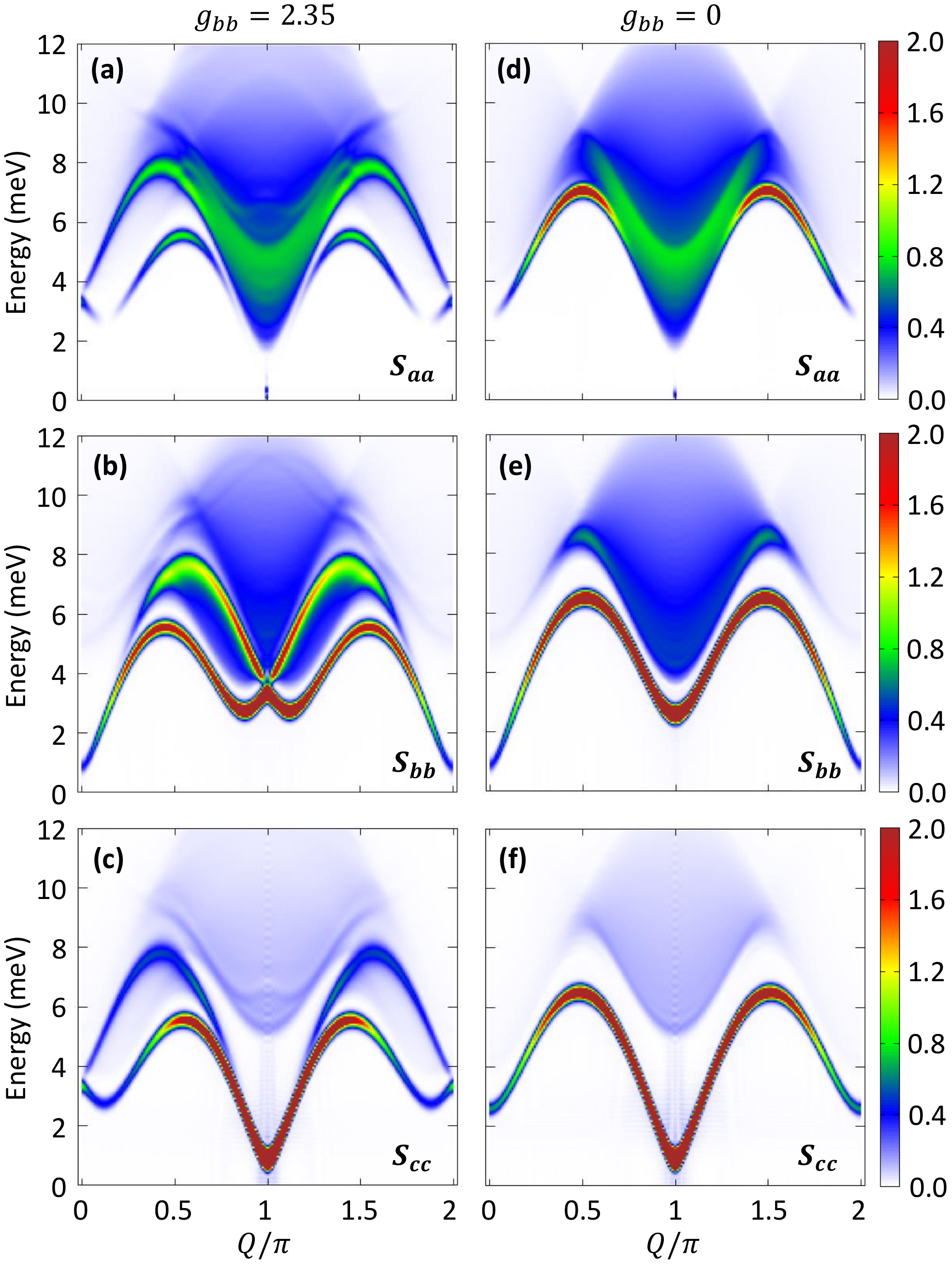}
\caption{\label{fig5} Dynamical susceptibility calculated for the simplified model of Eq.~\eqref{eq:SimpleHamil} describing a linear spin chain with $\epsilon=0.53$, $\mu_{0}H=12\;\mathrm{T}$.~$(g_{ba},g_{bb})$ are equal to (0.94,2.35) for panels (a-c) and equal to (0.94,0) for panels (d-f).~The components of dynamical susceptibility (a,d) $S_{aa}$, (b,e) $S_{bb}$, and (c,f) $S_{cc}$ are shown.}
\end{figure}

\begin{figure}
\centering
\includegraphics[width=7.5cm]{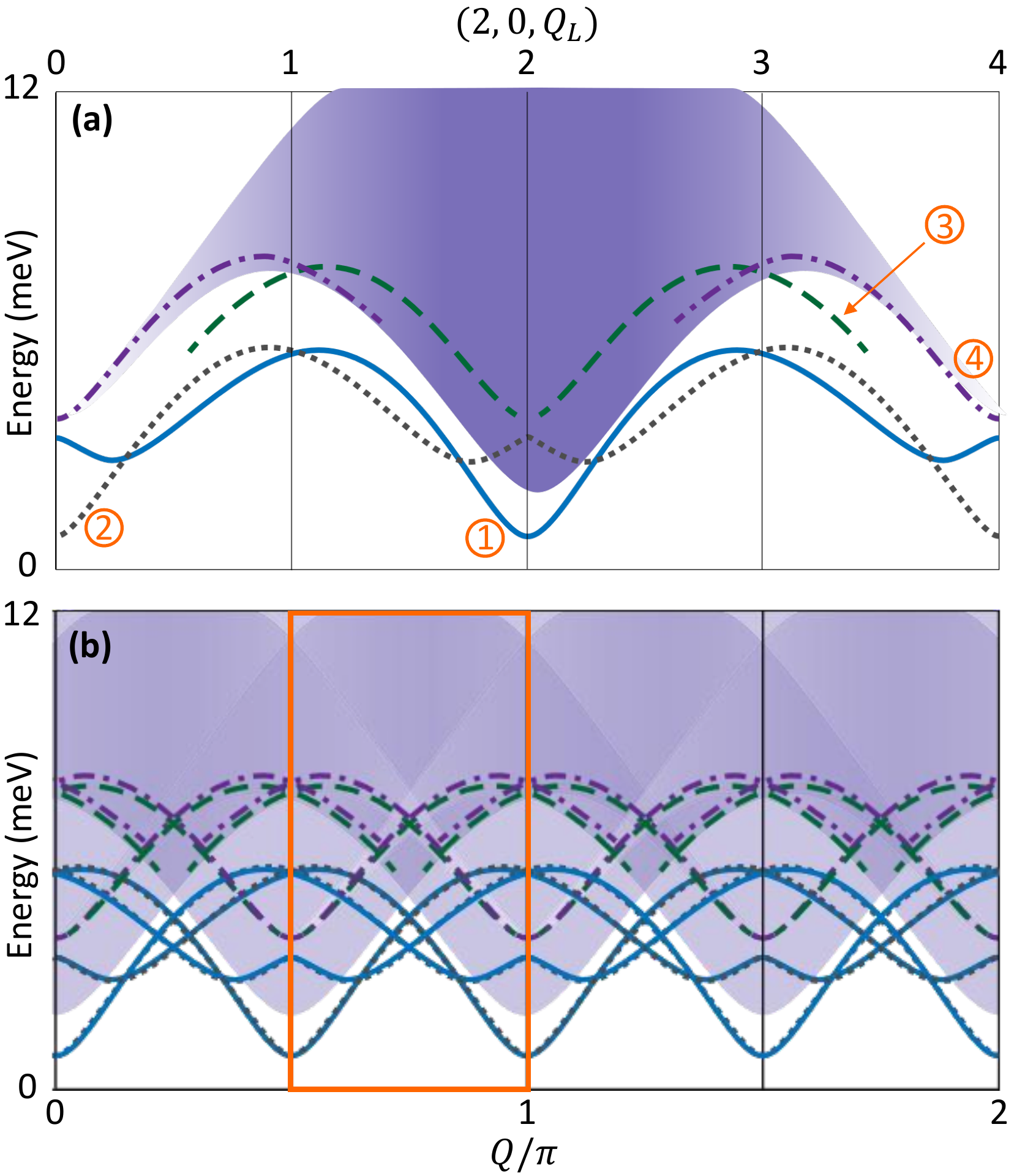}
\caption{\label{fig6} Schematic drawing of the excitation spectrum of \bcv\ with the same labeling of the modes as in Fig.~\ref{fig2} without (a) and with (b) the replica due to the presence of 4 ions in the lattice unit cell describing the screw chains (see main text, subsection III.C).}
\end{figure}

To illustrate this point, it is useful to consider the simplified model given by Eq.~\eqref{eq:SimpleHamil}, where the spins occupy the sites of such a simple linear spin chain.~Figures~\ref{fig5}(a-c) show the calculated dynamical structure factors $S_{aa}, S_{bb}$, and $S_{cc}$ for the $\tilde{g}$ tensor values $(g_{ba},g_{bb})=(0.94,2.35)$, which correspond to \bcv.~The spectrum includes a continuum along with dispersing modes, and a one to one correspondence can be easily done with the full results displayed in Figs.~\ref{fig2}(a) and \ref{fig2}(d).~Mode \textcircled{1} is visible in $S_{cc}$ [Fig.~\ref{fig5}(c)] while mode \textcircled{2} is visible in $S_{bb}$ [Fig.~\ref{fig5}(b)].~The continuum as well as the dispersions of the modes have been reproduced in Fig.~\ref{fig6}(a), using reduced wave vectors expressed both in terms of the actual reciprocal units $2\pi/c$ (upper scale) and in terms of $1/c_0$ (lower scale).~The minimum of the continuum occurs at the ``antiferromagnetic point'' $Q=\pi/c_{0}$, which translates into the reduced value $2 \times \frac{2\pi}{c}$ for the actual screw chains of \bcv.~Similar correspondences can be done for the dispersive features.~Interestingly, the latter shall be described over a period consisting in 4 reciprocal lattice units.

Figure \ref{fig6}(b) shows a sketch of the full spectrum, including the replicas, to be compared to the actual case of \bcv.~The portion of reciprocal space probed in Fig.~\ref{fig2} is highlighted by an orange rectangle.

This analysis shows that the lowest interlaced energy modes \textcircled{1} and \textcircled{2} are thus basically identical.~A noticeable difference between the experimental and calculated maps is the size of the gaps, which is different in the measurements at $Q_L=1$ and $Q_L=2$.~This indicates a dispersion along the ${\bf a}$ and ${\bf b}$ directions resulting from the interchain coupling that is not fully captured by the mean-field treatment in the calculations.~Interestingly, two higher energy modes \textcircled{3} and \textcircled{4} are also visible although less clearly, which also seem to be basically identical.

The spectral weight of the different modes also varies at various ``equivalent'' $Q_L$ positions.~This is related to both an interference effect due to the atomic positions but also to the fact that neutrons are sensitive only to the magnetic correlations between spin components perpendicular to the scattering vector.~This explains the difference in both maps of Figs.~\ref{fig2}(a) and \ref{fig2}(d), with in particular the absence of mode \textcircled{1} in the $(0,0,Q_L)$ map since, as discussed above, it is mainly polarized along the ${\bf c}$ direction.

\subsection{Shape of the dispersions}

It is worth noting that modes \textcircled{1} and \textcircled{2} present two kinds of minima, located at commensurate and incommensurate positions.~As shown in Fig.~\ref{fig2} and schematized in Fig.~\ref{fig6}, they are located at $Q_L=2$ (modulo 1) and at $Q_L\simeq1.74$ and $Q_L\simeq2.26$ (modulo 1), respectively.

Although the effect of the uniform magnetic field is relatively minor compared to the effect of the staggered one, it is nevertheless responsible for this incommensurate feature.~To prove this relationship, we performed numerical calculations using the simplified linear spin chain model Eq.~\eqref{eq:SimpleHamil}.~Figures~\ref{fig5}(d-f) show the calculations carried out for $(g_{ba},g_{bb})=(0.94,0)$ i.e.~without uniform field.~The comparison with Figs.~\ref{fig5}(b) and \ref{fig5}(e) demonstrates that the incommensurate features in the spectrum only appear in the presence of the external uniform magnetic field.~In addition, as shown in Figs.~\ref{fig5}(b) and \ref{fig5}(e) as well as Figs.~\ref{fig5}(c) and \ref{fig5}(f), it is clear that the uniform field produces the minima at incommensurate positions around $Q/\pi = 0$ for $S_{aa}$ and around $Q/\pi = 1$ for $S_{bb}$.~In Appendix~\ref{sec:bosonization}, we also show how this incommensurability can be understood from the viewpoint of the bosonized field theory \cite{Giamarchi2004}.

Furthermore, the  incommensurability shifts from $\pi$ to 0 in reduced momentum transfer with increasing magnetic field, as already mentioned by Matsuda {\it et al.}~\cite{Matsuda2017}.~In this spirit, numerical calculations were also performed at a higher magnetic field of 15.7 T, as shown in Figs.~\ref{fig4}(d-f).~With increasing field, the energy gaps increase at the minimum of the dispersion but also at the incommensurate positions, so that the whole excitation spectrum is pushed upward in energy.~The incommensurate minimum of modes \textcircled{1} and \textcircled{2} is also shifted with the field by approximately 0.06 r.l.u.~[see red dotted lines in Figs.~\ref{fig4}(b,e)].~The increase of the shift from the wave number $Q_{L}=\mathrm{integer}$ with increasing field can also be understood in the bosonized field theory since the $\nabla\phi(x)$ term (see Appendix~\ref{sec:bosonization}) becomes larger due to the increase of the uniform field along the {\bf b} axis.

\section{Discussion} \label{sec:discuss}

\bcv\ is a model system of the $XXZ$ spin chain antiferromagnet.~In zero field, its easy-axis anisotropy along the ${\bf c}$ axis forces the magnetic moments to lie along the ${\bf c}$ axis in a N\'eel order driven by the interchain coupling.~This is opposed by effective staggered and uniform magnetic field along the ${\bf a}$ and ${\bf b}$ axes respectively, both produced by the combination of a uniform magnetic field applied along ${\bf b}$ and of the nondiagonal $g$ tensor.~This competition between the antiferromagnetic orders along the ${\bf a}$ and ${\bf c}$ directions provokes a quantum phase transition at 10 T.~This transition is in the universality class of the transverse field Ising model~\cite{Pfeuty1970} as identified from the dual field double-sine Gordon model~\cite{faure2018,Takayoshi2018}.~However it is important to note that the dispersion of excitations in the two phases surrounding the transition is quite different from the ones of a transverse field Ising model.~Indeed, in the low field phase, the terms proportional to $\epsilon$ in the Hamiltonian lead to the dynamics of domain walls (spinons) even at zero magnetic field, i.e., the excitations have a momentum dependent dispersion differently from the transverse field Ising model (see Appendix~\ref{sec:trIsing}).

As emphasized in our previous works~\cite{faure2018,Takayoshi2018}, the spinons correspond to solitons of the $\phi$ bosonic field~\cite{Giamarchi2004} in the language of the bosonized field theory.~They are confined by the weak effective potential induced from the interchain interaction.~In contrast, in the high field phase, the new excitations are solitons of the $\theta$ bosonic field, which is conjugate to $\phi$.~The transition at $\mu_0H_c$ was thus characterized as a topological transition from the $\phi$-locked phase to the $\theta$-locked one.~More precisely, the excitations come into strongly bounded pairs of $\theta$-solitons, sitting on neighboring sites, and confined by the large staggered field in the ${\bf a}$ direction arising from the non-diagonal $g$ tensor.~The modes \textcircled{1} and \textcircled{2} discussed above are constructed on the basis of these pairs of $\theta$-solitons: as sketched in Fig.~\ref{fig1}(c), the hopping is actually accompanied by a flip from the $S_a=\pm1$ to the $S_a=\mp1$ index.~As a result, modes \textcircled{1} and \textcircled{2} are linear combinations of those solitons, forming transverse excitations with respect to the ${\bf a}$ axis, and polarized along the ${\bf b}$ or ${\bf c}$ axis.~It is worth noticing that, beyond these well defined modes, some continuum is still visible in the $S_{bb}$ and $S_{aa}$ components, hence pointing to the persistence of deconfined excitations which remain longitudinal to some extend.

It is thus also clear from this physical analysis that the interchain interactions become less relevant above $\mu_0H_c$.~At low field, the series of discrete energies is a direct manifestation of the linear potential between $\phi$-solitons induced by those interchain couplings.~Each mode corresponds to a typical average distance between solitons.~In contrast, in the high field phase, the interchain interactions play little role: not only is the magnetic structure dictated by the staggered field but this field also confines the $\theta$-solitons into bound pairs separated by one site [see Fig.~\ref{fig1}(c)].~As a result, the concept of  typical distance between solitons disappears above $\mu_0H_c$.~It remains that the interchain couplings likely need additional analysis: a more rigorous treatment beyond the mean field approximation may bring some new insight.~In the same spirit, a more detailed description involving different exchange paths along with anisotropic versus isotropic coupling constants could be interesting to investigate in future work.

\section{Conclusion} \label{sec:conclusion}

\bcv\ is a very rich material gathering, from its chemical and crystallographic architecture, many ingredients at the origin of remarkable behaviors in the field of quantum magnetism: Ising-like anisotropy, large intrachain versus weak but non-negligible interchain interactions, anisotropic $g$ tensor producing easy-axis anisotropy and effective staggered fields under the application of an external magnetic field.~Our combined experimental and numerical study of the full dispersive spectrum along the chain direction allows us to understand and pinpoint the role of each parameter, in particular of both uniform and staggered fields, in the spin dynamics when considering the problem of a magnetic field applied perpendicular to the chain axis.~It also provides a novel understanding of the properties of solitonic excitations, well beyond the canonical spinon continuum of the isotropic Heisenberg chain of spin 1/2.

\acknowledgments

We thank E.~Villard and B.~Vettard for their technical support during the inelastic neutron scattering measurements on ThALES and IN12, respectively, J.~Debray, A.~Hadj-Azzem, and J.~Balay for their contribution to the crystal growth, cut, and orientation.~We acknowledge ILL for allocation of neutron beamtime.~S.~T.~is supported by JSPS KAKENHI Grant No.~JP21K03412 and JST CREST Grant No.~JPMJCR19T3, Japan.~This work was supported in part by the Swiss National Science Foundation under Division II.~

\appendix

\section{Excitations and comparison with the transverse field Ising model}
\label{sec:trIsing}

As discussed already in Ref.~\cite{faure2018}, the main features of \bcv\ are described by the Hamiltonian Eq.~\eqref{eq:Hamil}.~Although this Hamiltonian itself is quite complicated, three main features are clearly important for describing its basic physics: (i) the large Ising anisotropy along the {\bf c} direction; (ii) the staggered transverse magnetic field along the {\bf a} axis, which is created by the staggered $g$ tensor and by the presence of the uniform magnetic field applied along the {\bf b} axis; (iii) the weak interchain coupling.

In order to analyze the physics of the system, it is convenient to first discuss the simplified Hamiltonian given by Eq.~\eqref{eq:SimpleHamil}.~In the first step, both the interchain coupling and the uniform magnetic field are ignored to focus on the effects of the staggered transverse field.~This leads to the strictly one-dimensional Hamiltonian:
\begin{align}
 {\cal H}_{\mathrm{stag}} =& J \sum_{n} [\epsilon ( S_{n}^a S_{n+1}^a + S_{n}^b S_{n+1}^b ) + S_{n}^c S_{n+1}^c] \nonumber\\
&- \tilde{h}_{a} \sum_{n} (-1)^n S^a_{n}
\label{eq:stag}
\end{align}
where $c$ corresponds to the Ising axis and $\tilde{h}_{a}$ is the staggered part of the magnetic field, in which the $g$ and $\mu_{\mathrm{B}}$ factors have been absorbed.

An even simpler version of this Hamiltonian would be to completely neglect the $\epsilon$ term leading to a purely Ising Hamiltonian along the chains:
\begin{align}
 {\cal H}_{\mathrm{ising}} = J \sum_{n} S_{n}^c S_{n+1}^c - \tilde{h}_{a} \sum_{n} (-1)^{n} S^a_{n}.
\label{eq:ising}
\end{align}
For this Hamiltonian, a simple gauge transformation $\tilde S_n^{a,b} = (-1)^n S_n^{a,b}$, $\tilde S_n^{c} = S_n^{c}$ removes the oscillating term and brings the problem back to the standard form of the transverse field Ising model:
\begin{equation}
 {\cal H}_{\mathrm{trans}} = J \sum_{n} \tilde{S}_{n}^c \tilde{S}_{n+1}^c - \tilde{h}_{a} \sum_{n} \tilde{S}^a_{n}.
\end{equation}
This Hamiltonian, which can be fully solved by a mapping onto free fermions~\cite{Pfeuty1970}, has a celebrated quantum phase transition separating two phases, one dominated by a staggered order along the $c$ direction and one with a uniform polarisation of the $\tilde{S}_n^a$ operator (and thus a staggered order for $S_n^a$).

One might thus naively think that the physics of \eqref{eq:stag} is simply that of the transverse field Ising model up to the gauge transformation.~This naive view, however, is not correct and although the quantum phase transition occurring in \eqref{eq:stag} is indeed in the universality class of the Ising model~\cite{faure2018,Takayoshi2018}, the nature of the phases, in particular the dispersion of the excitations, is strongly affected by the $S_{n}^{a} S_{n+1}^{a}$ and $S_{n}^{b} S_{n+1}^{b}$ terms present in Eq.~\eqref{eq:stag} and absent in Eq.~\eqref{eq:ising}.~For example, in absence of magnetic field, Eq.~\eqref{eq:stag} has a dispersion of the excitation due to the term
$S_{n}^{a} S_{n+1}^{a}+S_{n}^{b} S_{n+1}^{b} = \frac{1}{2}(S_{n}^{+} S_{n+1}^{-}+S_{n}^{-} S_{n+1}^{+})$, while an excitation created in the chain described by Eq.~\eqref{eq:ising} is unable to move.~Such a difference between the models is of course crucial when comparing with the neutron scattering spectra which clearly demonstrates the strongly dispersing nature of the excitations.

To illustrate this point, let us first carry out the gauge transformation to bring the staggered field into a uniform one, leading to
\begin{align}
 \mathcal{H}_{\mathrm{stag}}
=& J \sum_{n} \Big[- \epsilon \Big(\tilde{S}_{n}^a \tilde{S}_{n+1}^a + \tilde{S}_{n}^b \tilde{S}_{n+1}^b \Big) + S_{n}^c S_{n+1}^c\Big] \nonumber\\
&- \tilde{h}_{a} \sum_{n} \tilde{S}^a_{n}
\label{eq:stagtrans}
\end{align}
We then can use the Kramers-Wannier transformation~\cite{KramersWannier}
\begin{align}
 \tau_{n+\frac{1}{2}}^{b}
   =\prod_{p=1}^{n} \sigma_{p}^{a},\quad
 \tau_{n+\frac{1}{2}}^{c}
   =\sigma_{n}^{c}\sigma_{n+1}^{c},
\end{align}
to rewrite the Hamiltonian in terms of the bond operators, where $\sigma_{n}^{\alpha}$ and $\tau_{n}^{\alpha}$ are Pauli matrices, and $S_{n}^{\alpha}=\frac{1}{2}\sigma_{n}^{\alpha}$. Then the Hamiltonian is transformed into
\begin{align}
 {\cal H}_{\mathrm{stag}} =&
   \frac{J}{4} \sum_{n} \Big[\epsilon \Big(-1+\tau^c_{n+\frac{1}{2}}\Big)
     \tau^b_{n-\frac{1}{2}}\tau^b_{n+\frac{3}{2}}
     +\tau_{n+\frac{1}{2}}^{c}\Big] \nonumber\\
   &-\frac{\tilde{h}_{a}}{2}
     \sum_{n}\tau_{n-\frac{1}{2}}^{b}\tau_{n+\frac{1}{2}}^{b}.
\label{eq:bondham}
\end{align}
In this representation, the N\'eel order along the $c$ axis existing in the original chain for $\tilde{h}_{a} = 0$ corresponds to the uniform polarization along the $\tau^c = -1$ direction in the $\tau$ representation.~The excitations, which are the spinons of the original model, correspond to simply flipping a single spin in the latter.~When $\epsilon = \tilde{h}_{a} = 0$, such excitations would have no dynamics, as is the case of the transverse field Ising model.~Such dynamics only appears in the transverse field Ising model when $\tilde{h}_{a}$ is nonzero.~As can be seen from Eq.~\eqref{eq:bondham}, the $\tilde{h}_{a}$ term causes a hopping between two nearest neighbors and thus allows the spinon to hop by \textit{one} lattice site.~This gives rise to terms dispersing as $\cos(k a)$ in the excitation spectrum.

In contrast, the $\epsilon$ term in the $XXZ$ model leads to a quite different dispersion.~Indeed \textit{even} at $\tilde{h}_{a} = 0$, such terms allow the spinons to move.~In Eq.~\eqref{eq:bondham} for $\tilde{h}_{a} = 0$,
$\epsilon(-1+\tau^c_{n+\frac{1}{2}})\simeq \epsilon(-1+\langle\tau^c_{n+\frac{1}{2}}\rangle)\neq 0$, and the $\tau_{n-\frac{1}{2}}^{b}\tau_{n+\frac{3}{2}}^{b}$ term connects the site $n-\frac{1}{2}$ and $n+\frac{3}{2}$, which allows the spinon to move by \textit{two} lattice spacings, as can also be directly seen by other methods~\cite{Giamarchi2004}.~As a result, the modes have a dispersion which is essentially $\cos(2ka)$.~The excitations in the low field phase thus have a different structure and a different dispersion relation than the one predicted by the transverse field Ising model.~

In the high transverse field phase, the uniform field in Eq.~\eqref{eq:stagtrans} is dominant in the Hamiltonian and the magnetic order in the $+\tilde{S}_{n}^{a}$ direction grows.~In this sense, the situation is similar to the disordered phase of the transverse field Ising model.~However the uniform field along the $b$ axis, which is neglected in Eq.~\eqref{eq:stag} for simplicity, affects the dispersion, as we discussed in the main text.

Another illustration of this physics is provided by the field theory representation of the Hamiltonian Eq.~\eqref{eq:stag} established in Ref.~\cite{faure2018}.~The field theoretical approach for the high field phase is discussed in the next section.

\section{Bosonized field theory} \label{sec:bosonization}

We consider the simplified model Eq.~\eqref{eq:SimpleHamil}
with retaking the spin axes as
$S_{n}^{a}\to S_{n}^{a}$,
$S_{n}^{b}\to S_{n}^{c}$, and
$S_{n}^{c}\to -S_{n}^{b}$
($\pi/2$ rotation around the $a$ axis),
\begin{align}
 \mathcal{H} =&
  J \sum_{n} (
   \epsilon S_{n}^{a}S_{n+1}^{a}+S_{n}^{b}S_{n+1}^{b}
   +\epsilon S_{n}^{c}S_{n+1}^{c}] \nonumber\\
  &-g_{ba}\mu_{\mathrm{B}}\mu_{0}H\sum_{n}(-1)^{n}S_{n}^{a}
   -g_{bb}\mu_{\mathrm{B}}\mu_{0}H\sum_{n}S_{n}^{c}
\nonumber\\
  =&J \sum_{n} \frac{1+\epsilon}{2}\bigg[
   (S_{n}^{a}S_{n+1}^{a}+S_{n}^{b}S_{n+1}^{b})
   +\frac{2\epsilon}{1+\epsilon} S_{n}^{c}S_{n+1}^{c}\nonumber\\
   &-\frac{1-\epsilon}{1+\epsilon}
   (S_{n}^{a}S_{n+1}^{a}-S_{n}^{b}S_{n+1}^{b})\bigg] \nonumber\\
  &-g_{ba}\mu_{\mathrm{B}}\mu_{0}H\sum_{n}(-1)^{n}S_{n}^{a}
   -g_{bb}\mu_{\mathrm{B}}\mu_{0}H\sum_{n}S_{n}^{c}.
\end{align}
The bosonization formula for the spin operators are \cite{Giamarchi2004}
\begin{align}
\begin{split}
 S_{j}^{c}&\simeq -\frac{c_0}{\pi}\frac{d\phi(z)}{dz}
  +a_{1}(-1)^{j}\cos(2\phi(z)),\\
 S_{j}^{+}&\simeq e^{-i\theta(z)}
  [b_{0}(-1)^{j}+b_{1}\cos(2\phi(z))],
\end{split}
\label{eq:SpinBosonSM}
\end{align}
where $c_0$ is the lattice constant and $a_{1}$, $b_{0}$ and $b_{1}$ are some constants.~Conjugate bosonic fields $2\phi$ and $\theta$ can intuitively be considered as the polar and azimuthal angles of the N\'eel order.~The terms appearing in the Hamiltonian are bosonized following~\cite{Giamarchi2004,Takayoshi2018}
\begin{align}
 \sum_{n}(S_{n}^{a}S_{n+1}^{a}-S_{n}^{b}S_{n+1}^{b})
  &\simeq -\int dz\cos(2\theta(z))
\nonumber\\
 \sum_{n}S_{n}^{c}
  &\simeq -\frac{1}{\pi}\int dz \nabla\phi(z)
\nonumber\\
 \sum_{n}(-1)^{n}S_{n}^{a}
  &\simeq \int dz\cos\theta(z).
\nonumber
\end{align}
Thus the bosonized field theory is given as
\begin{align}
 \mathcal{H}_{\mathrm{bos}}
  &=\frac1{2\pi} \int dz
   \bigg[uK (\nabla \theta)^2
   +\frac{u}{K} (\nabla \phi)^{2} \bigg]
\nonumber\\
 &-g_{1} \int dz \cos( \theta(z))
  +g_{2} \int dz \cos(2\theta(z))
\nonumber\\
 &+g_{3} \int dz \cos(4 \phi(z))
  +g_{4} \int dz \nabla \phi(z)
\label{eq:HamilBosonSM}
\end{align}
where $u$ is the spinon velocity and $K$ the Tomonaga-Luttinger parameter. The $g_i$ are related to the original parameters in (B1): $g_1 \propto H$, $g_2 \propto \frac{J(1-\epsilon)}{1+\epsilon}$, $g_3 \propto \frac{2\epsilon J}{1+\epsilon}$ and $g_4 \propto H$. Since $2\epsilon/(1+\epsilon)<1$, the $\cos(4 \phi(z))$ term is irrelevant.~The $\cos(\theta(z))$ term is more relevant than the $\cos(2\theta(z))$ term.~Therefore the field $\theta(z)$ is locked at $2n\pi$ ($n$ is an integer).~From Eq.~\eqref{eq:SpinBosonSM}, the spin-spin correlation function $\langle S^{c}_{n}(t)S^{c}_{0}(0)\rangle$ is related to the correlation function of the $\phi(z)$ field.~Due to the existence of the $\nabla \phi(z)$ term in Eq.~\eqref{eq:HamilBosonSM}, the dispersion of $S_{cc}(q,\omega)$ excitations has an energy local minimum at some incommensurate momentum.

For the magic value $K=1/4$ this Hamiltonian can be refermionized \cite{Giamarchi2004} and mapped onto a free fermion model with both a staggered periodic potential (proportional to $g_3$) and a pairing term (proportional to $g_1$). Retaining only these two terms and diagonalizing the fermionic Hamiltonian by the usual Bogoliubov transformation leads for the energy of the lowest mode to 
\begin{equation}
 E_k = - \sqrt{\xi(k)^2 + (g_3-g_1)^2}
\end{equation}
with $\xi(k) = - J \epsilon \cos(k)$ and assuming $g_3>0$ and $g_1 >0$. The momentum $k$ is varying between $[0,\pi/2]$.~For $g_1=0$ one recovers the gap due to the Ising anisotropy. Expanding the square root gives back the spinon dispersion $\cos(2k)$. At the critical point $g_3=g_1$ the spectrum becomes massless and the duality of the model is apparent between $g_3 >  g_1$ and $g_3 < g_1$ but with an order for the pseudo-fermions going from an order in the density $g_3 < g_1$ to a pairing order $g_3 < g_1$.


\end{document}